\documentclass{article}

\usepackage[english]{babel}
\usepackage[utf8]{inputenc}
\usepackage{amsmath,amssymb}
\usepackage{parskip}
\usepackage{graphicx}

\usepackage[top=2.5cm, left=3cm, right=3cm, bottom=4.0cm]{geometry}
\usepackage[table]{xcolor}
\title{Spin-Torsion effect on collapsing of first generation stars into neutron stars rather than black holes in Einstein-Cartan-Sciama-Kibble theory}
\author{Emre Dil}
\date{\today}

\begin{document}
\maketitle
\begin{abstract}
    In this project, we try to find the correlation between the non-local pressure inside the massive neutron stars resisting the gravitational collapse of the core and ECSK dark energy led by the effect of spin-torsion coupling between quark fields and the space-time at very high densities much larger than the nuclear density. The injection of dark energy into the core of massive neutron stars (MANs) and extra resistant nature of this dark energy to the collapse of MANs by the anti-gravity give the possibility of existence of neutron stars in the unobserved mass range of $[2.16M_{\odot},5M_{\odot}]$. Obtaining the ECSK TOV equation gives the local pressure of the ambient medium of MANs. Moreover, the negative pressure from the ECSK dark energy is obtained from the Lagrangian again, from which we are able to investigate the hydro-static equilibrium of the core and ambient medium of the MANs. If the equilibrium state is satisfied for the unobserved mass gap for the MANs in ECSK theory framework this will imply our model predicts this vast mass range of unobserved spectrum of the MANs in astrophysical studies.
\end{abstract}
\section{Introduction}

The observed spectrum of neutron stars ranges between $[1.1M_{\odot},2.16M_{\odot}]$. There is a big unobserved mass range through the smallest mass of black hole with mass $5M_{\odot}$. It is possible to think that the old massive neutron stars (MANs) created from the first generation of stars just after the Big Bang may turn out to be the neutron stars in that unobserved mass range of $[2.16M_{\odot},5M_{\odot}]$. These MANs are considered as completely invisible objects with incompressible quark-gluon superfluid and very high central density much larger than the nuclear one \cite{1,2,3}.

In the literature dark energy injection mechanism is proposed to explain the existence of those invisible MANs \cite{1}. Their quark-gluon cores are stratified by a scalar field leading to the increase of non-local pressure which resisting the gravitational collapse of the core. In this study we will propose a completely different mechanism than the injection of dark energy, which results from the spin-torsion coupling between the space-time of ambient medium of outer core of the neutron star and the high density quark matter field in this ambient medium in the framework of Einstein-Cartan-Sciama-Kibble (ECSK) theory.

ECSK theory is a generalization of Einstein's general relativity which involves the matter field with half-integer spin \cite{4,5,6}. The inclusion of this spinor field leads to a space-time with torsion in addition to the curvature. In this study, we consider these spin-torsion effects of quark fields and the space-time of ambient medium of neutron star, residing just out of the flat core region. In previous studies, this ambient medium has been considered as curved space-time \cite{1,3}, however we do not consider this ambient medium only curved but also twisted due to the torsion effect of spinor quark field in ECSK theory framework.

In cosmological studies, the effect of spin-torsion coupling between quark fields and the space-time at very high densities much larger than the nuclear density acts as dark energy due to the four-fermion interaction terms in the Dirac Lagrangian in ECSK theory \cite{4}. Because the injection of dark energy is assumed to be responsible from the non-local pressure inside the massive neutron stars resisting the gravitational collapse of the core in the general relativistic framework in previous studies, we generalize this point of view by considering the quark fields and the ambient medium interactions creating the dark energy effect due to the four-fermion interaction terms in ECSK theory framework. We will investigate the possibility of existence of neutron stars in the unobserved mass range of $[2.16M_{\odot},5M_{\odot}]$ by assuming the injection of gravitationally resistant dark energy into the core of MANs, due to the interaction of spinor quark fields and the twisted ambient medium with space-time torsion.

\section{Massive Neutron Stars in ECSK theory}

Massive neutron stars (MANs) are observed to get spin down resulting from the loss of rotational and magnetic energies with a ratio of $10^{38}erg/s$, which means that the exhaustion of total removable energies take about ten million years by the realization of higher heat conductivity operating on larger length scales than the nuclear scales. Accordingly, very old MANs formed by the collapse of first generation stars must be dark and invisible due to the loss of all removable energies and this also gives explanation why there is an observational mass gap between $[2.16M_{\odot},5M_{\odot}]$ for neutron stars \cite{3}.

These loss of energies due to the collapse of progenitors of MANs are transported out toward the ambient medium of the core of the stars. Although the nuclear energy generation stops, the Tolman-Oppenheimer-Volkoff (TOV) equation still allows a thermal energy transfer from core to the ambient medium by which the core of neutron star enters in a completely incompressible superfluid state  from a dissipative compressible phase \cite{3}. We will investigate these mechanism in the ECSK theory framework by considering the spin-torsion couplings in the space-times of super-baryon core region formed by the merger of neutrons to a quark-gluon cloud and the baryonic ambient medium.

\subsection{ECSK formalism}

The dynamics of spinor fields, such as quarks, leptons and all kind of fermions in curved space-times can be investigated by using the ECSK theory formalism. Obtaining the energy-momentum tensor of Dirac spinors in curved space-times is a complex work, because the spinor field cannot be held constant during the variation of the metric as the Lagrangian varies with the variations of the metric. For spinor fields, when the metric varies, the spinor field components also vary and they cannot be held constant with respect to some fixed frame as in the scalar field case. Therefore, we have to give the algebraic structure of the ECSK formalism and the spinor field structures.

The dynamical variables of metric-affine ECSK formulation of the gravity are the tetrad frame field $e_{a}^{i} $ and the spin connection $\omega _{bk}^{a} =e_{j}^{a} e_{b;k}^{j} =e_{j}^{a} (e_{b,k}^{j} +\Gamma _{ik}^{j} e_{b}^{i} )$, comma denoting the ordinary partial derivative with respect to coordinate $x^{k}$, and semicolon denoting the covariant derivative with respect to the affine connection $\Gamma _{jk}^{i} $ whose antisymmetric lower indices give the torsion tensor $S_{jk}^{i} =\Gamma _{[jk]}^{i} $. The tetrad fiels relates the space-time coordinates indices $i,j,...$ and the local Lorentz coordinates indices $a,b,...$, by $V^{a} =V^{i} e_{i}^{a} $ where $V^{a} $ and  $V^{i} $ are a Lorentz vector and a standard vector, respectively. The covariant derivative of Lorentz vectors is  denoted by a bar and defined in terms of the spin connection: $V_{\left|i\right. }^{a} =V_{,i}^{a} +\omega _{bi}^{a} \, V^{b} $ and $V_{a\left|i\right. } =V_{a,i} -\omega _{ai}^{b} \, V_{b} $, and the semicolon covariant derivative of standard vectors is defined in terms of the affine connection, $V_{;i}^{k} =V_{,i}^{k} +\Gamma _{li}^{k} \, V^{l} $ and $V_{k;i} =V_{k,i} -\Gamma _{ki}^{l} \, V_{l} $. One can lower-raise Local Lorentz coordinates by the Minkowski metric $\eta _{ab} $, and the space-time coordinates by the metric tensor $g_{ik} $. The metricity condition $g_{i\, j;k} =0$ gives the definition of affine connection as $\Gamma _{i\, j}^{k} =\{ _{i\, j}^{k} \} +C_{i\, j}^{k} $ in terms of the Christoffel symbols $\{ _{i\, j}^{k} \} =(1/2)g^{km} (g_{mi,j} +g_{m\, j,i} -g_{i\, j,m} )$ and the contortion tensor $C_{jk}^{i} =S_{jk}^{i} +2S_{(jk)}^{i} $. The symmetrization reads $A_{(jk)} =(1/2)(A_{jk} +A_{k\, j} )$, and the antisymmetrization by $A_{[jk]} =(1/2)(A_{jk} -A_{k\, j} )$ throughout the paper. In ECSK formalism, we can also take the dynamical variables as the metric $g_{ik} =\eta _{ab} e_{i}^{a} e_{k}^{b} $ and the torsion $S_{ik}^{j} =\omega _{[ik]}^{j} +e_{[i,k]}^{a} e_{a}^{j} $ rather than the tetrad frames and the spin connection \cite{7}-\cite{25}. 

A tensor density $T_{D} $ can be related to the corresponding tensor $T$ by $T_{D} =eT$, where $e=\det e_{i}^{a} =\sqrt{-\det g_{ik} } $, therefore, the spin and the energy-momentum densities are given by $\sigma _{i\, jk} =e\, s_{i\, jk} $ and ${\rm T} _{ik} =eT_{ik} $. These are called as \textit{metric} spin tensor and energy-momentum tensor because they are specified by the space-time coordinate indices, and they are obtained from the variation of the Lagrangian with respect to the torsion (or contortion) tensor $C_{k}^{i\, j} $ and the metric tensor $g^{i\, j} $, respectively. Accordingly, the metric spin tensor is found to be $s_{i\, j}^{k} =(2/e)(n \ell _{m} /\delta C_{k}^{i\, j} )=(2/e)(\partial \ell _{m} /\partial C_{k}^{i\, j} )$, while the metric energy-momentum tensor is $T_{i\, j} =(2/e)(\delta \ell _{m} /\delta g^{i\, j} )=(2/e)[\partial \ell _{m} /\partial g^{i\, j} -\partial _{k} (\partial \ell _{m} /\partial (g_{,k}^{i\, j} ))]$. The Lagrangian density of the spinor field is $\ell _{m} =eL_{m} $. If the local Lorentz coordinates are used to specify these tensors as $\sigma _{ab}^{i} =e\, s_{ab}^{i} $ and ${\rm T} _{i}^{a} =eT_{i}^{a} $, then $s_{ab}^{i} $ and $T_{i}^{a} $ are called as the \textit{dynamical} spin tensor and \textit{dynamical} energy-momentum tensor, respectively, and they are obtained from the variation of the Lagrangian with respect to the tetrad $e_{a}^{i} $ and the spin connection $\omega _{i}^{ab} $, such that $s_{ab}^{i\, } =(2/e)(\delta \ell _{m} /\delta \omega _{i}^{ab} )=(2/e)(\partial \ell _{m} /\partial \omega _{i}^{ab} )$, and $T_{i}^{a} =(1/e)(\delta \ell _{m} /\delta e_{a}^{i} )=(1/e)[\partial \ell _{m} /\partial e_{a}^{i} -\partial _{j} (\partial \ell _{m} /\partial (e_{a,j}^{i} ))]$.

Total action of the gravitational field with spinor field in metric-affine ECSK theory is given in the same form with the classical Einstein-Hilbert action, such as $S=\kappa \int (\ell _{g} +\ell _{\psi } )d^{4} x $, where $\kappa =8\pi \, G$ and $\ell _{g} =-(1/2\kappa )eR$, and $\ell _{\psi } $ are the gravitational, and fermionic matter Lagrangian densities. The Ricci scalar is $R=R_{j}^{b} e_{b}^{j} $ where $R_{j}^{b} =R_{jk}^{bc} e_{c}^{k} $ is the Ricci tensor obtained from the curvature tensor $R_{jk}^{bc} $. Also, the curvature tensor is related to the spin connection, such that $R_{bi\, j}^{a} =\omega _{bj,i}^{a} -\omega _{bi,j}^{a} +\omega _{ci}^{a} \, \omega _{bj}^{c} -\omega _{cj}^{a} \, \omega _{bi}^{c} $. The variation of the action with respect to the contortion tensor gives the Cartan equations $S_{i\, k}^{j} -S_{i} \, \delta _{k}^{j} +S_{k} \, \delta _{i}^{j} =-(\kappa /2e)\, \sigma _{ik}^{j} $, and the variation with respect to the metric tensor yields the Einstein equations $G_{ik} =\kappa (T_{ik}^{\psi } +U_{ik}^{\psi } )$, where $G_{ik} =P_{i\, jk}^{j} -(1/2)P_{lm}^{lm} g_{ik} $ is the Einstein tensor. Here $P_{i\, jk}^{j} $ is also the Riemann curvature tensor given by the relation $R_{klm}^{i} =P_{klm}^{i} +C_{km\, :l}^{i} -C_{kl\, :\, m}^{i} +C_{km}^{j} C_{jl}^{i} -C_{kl}^{j} C_{jm}^{i} $, where colon represents the Riemannian covariant derivative with respect to the Levi-Civita connection $\{ _{li}^{k} \} \, $, such as $V_{:i}^{k} =V_{,i}^{k} +\{ _{li}^{k} \} \, V^{l} $ and $V_{k\, :i} =V_{k,i} -\{ _{ki}^{l} \} \, V_{l} $. Also, the curvature tensor transforms into the Riemann tensor for torsion-free general relativity theory. The $U_{ik} $ term in the Einstein equations is the spin contribution $U_{ik} =\kappa (-s_{[l}^{ij} s_{j]}^{kl} -(1/2)s^{i\, jl} s_{jl}^{k} +(1/4)s^{jli} s_{jl}^{k} +(1/8)g^{ik} (-4s_{j[m}^{l} s_{l]}^{jm} +s^{jlm} s_{jlm} ))$, and the total energy-momentum tensor of the spinor field is given by $\Theta _{ik}^{\psi } =T_{ik}^{\psi } +U_{ik}^{\psi } $.

Combining the above algebraic relations for the metric-affine ECSK formulation of gravity for (+,-,-,-) metric signature, spinor field is described by the Lagrangian densities of the form:
\begin{equation} \label{GrindEQ__1_} 
\ell _{\psi } =e(i/2)(\bar{\psi }\gamma ^{k} \psi _{;k} -\bar{\psi }_{;k} \gamma ^{k} \psi )-em_{\psi } \bar{\psi }\psi ,     
\end{equation} 
where $\psi $ and $\bar{\psi }=\psi ^{+} \gamma ^{0} $ are the spinor and the adjoint spinor fields, respectively. The semicolon covariant derivative of the spinor and the adjoint spinor fields are given by 
\begin{equation} \label{GrindEQ__2_} 
\psi _{;\, k} =\psi _{,\, k} -\Gamma _{k} \psi ,  
\end{equation} 
\begin{equation} \label{GrindEQ__3_} 
\bar{\psi }_{;\, k} =\bar{\psi }_{,\, k} -\Gamma _{k} \bar{\psi },      
\end{equation} 
where $\Gamma _{k} =-(1/4)\, \omega _{abk} \gamma ^{a} \gamma ^{b} $ is the Fock-Ivanenko spin connection, $\gamma ^{k} $ and $\gamma ^{a} $ are the \textit{metric} and \textit{dynamical} Dirac gamma matrices as; $\gamma ^{k} =e_{a}^{k} \gamma ^{a} $, $\gamma ^{(k} \gamma ^{m)} =g^{km} I$ and $\gamma ^{(a} \gamma ^{b)} =\eta ^{ab} I$. One can decompose the semicolon covariant derivative of the spinor field into a colon Riemannian covariant derivative with the contortion tensor $C_{i\, jk} $ term as
\begin{equation} \label{GrindEQ__4_} 
\psi _{;\, k} =\psi _{:\, k} +(1/4)C_{i\, jk} \gamma ^{[i} \gamma ^{j]} \psi ,           
\end{equation} 
\begin{equation} \label{GrindEQ__5_} 
\bar{\psi }_{;\, k} =\bar{\psi }_{:\, k} -(1/4)C_{i\, jk} \bar{\psi }\gamma ^{[i} \gamma ^{j]} .           
\end{equation} 
The colon Riemannian covariant derivative is also defined to be 
\begin{equation} \label{GrindEQ__6_} 
\psi _{:\, k} =\psi _{,\, k} +(1/4)\, g_{ik} \{ _{jm}^{i} \} \gamma ^{j} \gamma ^{m} \psi ,             
\end{equation} 
\begin{equation} \label{GrindEQ__7_} 
\bar{\psi }_{:\, k} =\bar{\psi }_{,\, k} -(1/4)\, g_{ik} \{ _{jm}^{i} \} \gamma ^{j} \gamma ^{m} \bar{\psi }.             
\end{equation} 
Although the spinor field Lagrange density contains covariant derivatives including the contortion tensor $C_{i\, jk} $, the explicit form of the contortion tensor is obtained from the Cartan equations whose right hand side involves the spin tensor density. Then, the spin tensor is led by the variation of the spinor Lagrangian with respect to the contortion tensor, such as
\begin{equation} \label{GrindEQ__8_} 
s^{i\, jk} =(1/e)\sigma ^{i\, \, jk} =-(1/e)\, \varepsilon ^{i\, jkl} s_{l} ,            
\end{equation} 
where $\varepsilon ^{i\, jkl} $ is the Levi-Civita symbol, and
\begin{equation} \label{GrindEQ__9_} 
s^{i} =(1/2)\, \bar{\psi }\gamma ^{i} \gamma ^{5} \psi  
\end{equation} 
is the spin pseudo-vector, and $\gamma ^{5} =i\gamma ^{0} \gamma ^{1} \gamma ^{2} \gamma ^{3} $. Substituting the spin tensor of spinor field in the Cartan equations leads to the torsion tensor as
\begin{equation} \label{GrindEQ__10_} 
S_{i\, jk} =C_{i\, jk} =(1/2)\kappa \varepsilon _{i\, jkl} s^{l} ,     
\end{equation} 
which will be found in the spinor field Lagrange density. 

Then, the variation of the spinor fermionic matter Lagrangian density with respect to the adjoint spinor $(\partial \ell _{\psi } /\partial \bar{\psi })-(\partial \ell _{\psi } /\partial \bar{\psi }_{:\, k} )_{:\, k} =0$ gives the ECSK Dirac equation
\begin{equation} \label{GrindEQ__11_} 
i\gamma ^{k} \psi _{:k} -m_{\psi } \psi +\frac{3}{8} \kappa (\bar{\psi }\gamma ^{k} \gamma ^{5} \psi )\gamma _{k} \gamma ^{5} \psi =0,    
\end{equation} 
while the variation with respect to the spinor itself $(\partial \ell _{\psi } /\partial \psi )-(\partial \ell _{\psi } /\partial \psi _{:\, k} )_{:\, k} =0$ gives adjoint ECSK Dirac equation as
\begin{equation} \label{GrindEQ__12_} 
i\bar{\psi }_{:k} \gamma ^{k} +m_{\psi } \bar{\psi }-\frac{3}{8} \kappa (\bar{\psi }\gamma ^{k} \gamma ^{5} \psi )\, \bar{\psi }\gamma _{k} \gamma ^{5} =0.    
\end{equation} 

\subsection{Existence of dark energy in ECSK theory}

The equation (11) and its adjoint conjugate (12) can also be obtained directly by varying $\bar{\psi}$ and $\psi$, for the following Lagrangian density, respectively \cite{22}:

\begin{equation}
\label{13}
\ell _{\psi } =\frac{i \sqrt{-g}}{2}\left(\bar{\psi} \gamma^{i} \psi_{: i}-\bar{\psi}_{: i} \gamma^{i} \psi\right)-m \sqrt{-g} \bar{\psi} \psi+\frac{3 \kappa \sqrt{-g}}{16}\left(\bar{\psi} \gamma_{k} \gamma^{5} \psi\right)\left(\bar{\psi} \gamma^{k} \gamma^{5} \psi\right)   
\end{equation}

where we don not vary it with respect to the torsion. By using the identity $\frac{\delta \gamma^{j}}{\delta g^{i k}}=\frac{1}{2} \delta_{(i}^{j} \gamma_{k)}$, the corresponding energy-momentum tensor $T_{i k}=$ $\frac{2}{\sqrt{-g}} \frac{\delta \ell _{\psi }}{\delta g^{i k}}$ is, given by:

\begin{equation}
\label{14}
\begin{aligned}
T_{i k}=& \frac{i}{2}\left(\bar{\psi} \delta_{(i}^{j} \gamma_{k)} \psi_{: j}-\bar{\psi}_{: j} \delta_{(i}^{j} \gamma_{k)} \psi\right)-\frac{i}{2}\left(\bar{\psi} \gamma^{j} \psi_{: j}-\bar{\psi}_{: j} \gamma^{j} \psi\right) g_{i k}+m \bar{\psi} \psi g_{i k} \\
&-\frac{3 \kappa}{16}\left(\bar{\psi} \gamma_{j} \gamma^{5} \psi\right)\left(\bar{\psi} \gamma^{j} \gamma^{5} \psi\right) g_{i k}
\end{aligned}  
\end{equation}

where the definition of the Dirac matrices
$\gamma^{(i} \gamma^{k)}=g^{i k} I$ is used.  Inserting (11) into (14) leads

\begin{equation}
    \label{15}
    T_{i k}=\frac{i}{2}\left(\bar{\psi} \delta_{(i}^{j} \gamma_{k)} \psi_{: j}-\bar{\psi}_{: j} \delta_{(i}^{j} \gamma_{k)} \psi\right)+\frac{3 \kappa}{16}\left(\bar{\psi} \gamma_{j} \gamma^{5} \psi\right)\left(\bar{\psi} \gamma^{j} \gamma^{5} \psi\right) g_{i k}
\end{equation}

The first term on the right hand side of (15) is the energy-momentum tensor of a spinor field without torsion, and the second term is the dark energy term in terms of an effective cosmological constant, such as \cite{11,25},

\begin{equation}
    \label{16}
    \Lambda=\frac{3 \kappa^{2}}{16}\left(\bar{\psi} \gamma_{j} \gamma^{5} \psi\right)\left(\bar{\psi} \gamma^{j} \gamma^{5} \psi\right)
\end{equation}

with the dark energy density,

\begin{equation}
    \label{17}
    \rho_{\Lambda}=\frac{3 \kappa}{16}\left(\bar{\psi} \gamma_{j} \gamma^{5} \psi\right)\left(\bar{\psi} \gamma^{j} \gamma^{5} \psi\right)
\end{equation}

This torsion resulted cosmological constant becomes a time varying quantity because it depends on spinor fields, however, it is still constant in space at cosmological scales in a homogeneous and isotropic universe.

The fermions of quarks and leptons are all ingredients of stars, and described by the Dirac equation in quantum field theory. Because the Dirac spinor fields couple minimally to the torsion tensor of the medium, it does not vanish at microscopic scales in the
presence of fermions \cite{13}. At macroscopic scales, these spinor particles are averaged as a Weyssenhoff spin fluid \cite{26,27} and if
orientation of the spins are random their average  vanishes. However, the quadratic terms in the spin tensor do not
vanish \cite{24} if the spinors are significant at higher densities than the density of nuclear matter.

The averaged macroscopic canonical energy-momentum tensor of a spin fluid whose microscopic description given above reads as

\begin{equation}
\label{18}
    \sigma_{i j}=c \Pi_{i} u_{j}-p\left(g_{i j}-u_{i} u_{j}\right)
\end{equation}

while its macroscopic canonical spin tensor reads

\begin{equation}
    \label{19}
    s_{i j}^{k}=s_{i j} u^{k}, \quad s_{i j} u^{j}=0
\end{equation}

where $\Pi_{i}$ is the four-momentum density of the fluid, $u^{i}$ its four velocity, $s_{i j}$ its spin density, and $p$ its pressure \cite{24}. Using (18), (19) and Einstein's equation given above for $G^{i j}$ we obtain

\begin{equation}
\label{20}
\begin{aligned}
G^{i j}=& \kappa\left(\rho-\frac{1}{4} \kappa s^{2}\right) u^{i} u^{j}-\kappa\left(p-\frac{1}{4} \kappa s^{2}\right)\left(g^{i j}-u^{i} u^{j}\right) \\
&-\frac{1}{2} \kappa\left(\delta_{k}^{l}+u_{k} u^{l}\right) \nabla_{l}^{|}\left(s^{k i} u^{j}+s^{k j} u^{i}\right)
\end{aligned}
\end{equation}

where $\rho=c \Pi_{i} u^{i}$ is the rest energy density of the fluid,

\begin{equation}
    \label{21}
    s^{2}=\frac{1}{2} s_{i j} s^{i j}>0
\end{equation}

is the square of the spin density, and $\nabla_{k}^{|}$ denotes the general relativistic covariant derivative with respect to the Christoffel symbols $\left\{i^k {} _{j}\right\}$. If the spin orientation of particles in a spin fluid is random then the last term on the right of (20) vanishes after averaging. Thus the Einstein-Cartan equations for such a spin fluid are equivalent to the Einstein equations for a perfect fluid with the effective energy density $\rho-\kappa s^{2} / 4$ and the effective pressure $p-\kappa s^{2} / 4$ [18,25-27]. Also by using (8),(9) and (21), we obtain

\begin{equation}
    \label{22}
    \rho_{\Lambda}=-\frac{1}{4} \kappa s^{2}=\frac{3 \kappa}{16}\left(\bar{\psi} \gamma_{j} \gamma^{5} \psi\right)\left(\bar{\psi} \gamma^{j} \gamma^{5} \psi\right)
\end{equation}

The average particle number density in a fluid, $n,$ is related to the energy density and pressure of the fluid by $d n / n=d \rho /(\rho+p)$ If the fluid is described by a barotropic equation of state $p=w \rho$ then $n \propto \rho^{1 /(1+w)} .$ The square of the spin density for a fluid consisting of fermions with no spin polarization is given by

\begin{equation}
    \label{22a}
    s^{2}=\frac{1}{8}(\hbar c n)^{2}
\end{equation}

which yields

\begin{equation}
    \label{22b}
    s^{2} \propto \rho^{2 /(1+w)}[26] .
\end{equation}

Substituting this relation into (12) gives [25]

\begin{equation}
    \label{22c}
    \rho \propto a^{-3(1+w)}
\end{equation}

which has the same form as for $s^{2}=0$ (in the absence of spin). Accordingly, the spin-density contribution to the total effective energy density in (10) scales like

\begin{equation}
    \label{22d}
    \rho_{S}=\rho_{\Lambda}=-\frac{1}{4} \kappa s^{2} \propto a^{-6}.
\end{equation}

\subsection{Path to find TOV equation in ECSK theory}

As in the torsion-free general relativity, we can obtain the TOV equation of hydro-static equilibrium by using the field equations for of a superfluid matter. Because we consider a spin fluid causing space-time torsion, we should use the field equations of ECSK theory, such as Cartan equations 
\begin{equation}
\label{23}
   S_{i\, k}^{j} -S_{i} \, \delta _{k}^{j} +S_{k} \, \delta _{i}^{j} =-(\kappa /2e)\, \sigma _{ik}^{j} 
\end{equation}
and the standard Einstein equations 
\begin{equation}
    \label{24}
    G_{ik} =P_{i\, jk}^{j} -\frac{1}{2}P_{lm}^{lm} g_{ik}=\kappa (T_{ik} +U_{ik} ) 
\end{equation}
This equation set is much more complex than the torsion-free general relativistic equation in order to obtain the TOV equations for a spin-fluid whose spin tensor density $\sigma _{ik}^{j} $, energy-momentum tensor $T_{ik}$ and spin contribution of energy-momentum tensor $U_{ik}$ can be obtained from the variations of the matter Lagrangian, such as 

\begin{equation}
    \label{25}
    \sigma_{i\, j}^{k} =2\frac{\partial \ell _{m}}{\partial C_{k}^{i\, j}}
\end{equation}

\begin{equation}
    \label{26}
    T_{i\, k} =\frac{2}{e}\left[\frac{\partial \ell _{m}}{\partial g^{i\, k}} -\partial _{j} \frac{\partial \ell _{m} }{\partial (g_{,j}^{i\, k} )}\right]
\end{equation}

\begin{equation}
    \label{27}
    U_{ik} =\kappa (-s_{[l}^{ij} s_{j]}^{kl} -\frac{1}{2}s^{i\, jl} s_{jl}^{k} +\frac{1}{4}s^{jli} s_{jl}^{k} +\frac{1}{8}g^{ik} (-4s_{j[m}^{l} s_{l]}^{jm} +s^{jlm} s_{jlm} ))
\end{equation}
where $s_{i\, jk}=\sigma _{i\, jk} /e\,  $ spin tensor can also be obtained from equation (\ref{25}). Here, the left hand sides of equations (\ref{23}) and (\ref{24}) contain the torsion tensor $S_{i\, k}^{j}$ and curvature tensors $P^j_{ijk}$, and they are both obtained through the metric tensor. By considering a spherically symmetric metric and spin superfluid for the ambient medium of the neutron star, we construct a complete system to obtain the TOV equations in ECSK theory, as in the general relativity case. The most important part of obtaining TOV equations is to set the spin-fluid quark matter Lagrangian of the neutron stars. 

\section{Purpose}

In this project, we try to find the correlation between the non-local pressure inside the massive neutron stars resisting the gravitational collapse of the core and ECSK dark energy led by the effect of spin-torsion coupling between quark fields and the space-time at very high densities much larger than the nuclear density. The injection of dark energy into the core of MANs and extra resistant nature of this dark energy to the collapse of MANs by the anti-gravity give the possibility of existence of neutron stars in the unobserved mass range of $[2.16M_{\odot},5M_{\odot}]$.

Although all the removable energies are transported out of the core region, TOV equation still allows a thermal energy transfer from core to the ambient medium. Therefore, the core of MAN enters in an incompressible superfluid phase  from a compressible dissipative phase \cite{3}. This implies that we need to obtain the ECSK version of TOV equation. The most important point in finding the TOV equation is to construct the suitable ECSK Lagrangian for the quark spin fluid and the twisted and curved ambient medium of the MANs.

Obtaining the ECSK TOV equation gives the local pressure of the ambient medium of MANs. Moreover, the negative pressure from the ECSK dark energy is obtained from the Lagrangian again, from which we are able to investigate the hydro-static equilibrium of the core and ambient medium of the MANs. If the equilibrium state is satisfied for the unobserved mass gap for the MANs in ECSK theory framework this will imply our model predicts this vast mass range of unobserved spectrum of the MANs in astrophysical studies.

\end{document}